\documentclass[10pt,twocolumn,letterpaper]{article}

\usepackage[pagenumbers]{cvpr} 
\usepackage{natbib}

\usepackage{xcolor}
\usepackage{verbatim}
\definecolor{cvprblue}{rgb}{0.21,0.49,0.74}
\usepackage[pagebackref,breaklinks,colorlinks,allcolors=cvprblue]{hyperref}

\usepackage{import}
\newcommand{\name}{RFScape\xspace}

\def\paperID{4948} 
\def\confName{CVPR}
\def\confYear{2025}

\title{ Radio Frequency Ray Tracing with Neural Object  Representation 
}

\author{Xingyu Chen\\
UC San Diego\\
{\tt\small xic063@ucsd.edu}
\and
Zihao Feng\\
UC San Diego\\
{\tt\small zif004@ucsd.edu }
\and
Kun Qian\\
University of Virginia\\
{\tt\small kunqian@virginia.edu  }
\and
Xinyu Zhang\\
UC San Diego\\
{\tt\small xyzhang@ucsd.edu }
}

\begin{document}
\maketitle



\begin{abstract}
    Radio frequency (RF) propagation modeling poses unique electromagnetic simulation challenges. While recent neural representations have shown success in visible spectrum rendering, the fundamentally different scales and physics of RF signals require novel modeling paradigms. In this paper, we introduce \name, a novel framework that bridges the gap between neural scene representation and RF propagation modeling. Our key insight is that complex RF-object interactions can be captured through object-centric neural representations while preserving the composability of traditional ray tracing. Unlike previous approaches that either rely on crude geometric approximations or require dense spatial sampling of entire scenes, \name learns per-object electromagnetic properties and enables flexible scene composition. Through extensive evaluation on real-world RF testbeds, we demonstrate that our approach achieves 13 dB improvement over conventional ray tracing and 5 dB over state-of-the-art neural baselines in modeling accuracy, while requiring only sparse training samples. 
   
\end{abstract}

\section{Introduction}
\label{sec:intro}

Recent advances in neural rendering have transformed 3D scene understanding and reconstruction in computer vision \cite{thies2019deferred, Xiang_2021_CVPR, chen2021mvsnerf, zhang2022nerfusion, Suhail_2022_CVPR, lombardi2021mixture, Yu_2021_ICCV, Li_2023_CVPR, lindell2021autoint, yang2023freenerf}. Methods such as Neural Radiance Fields \cite{mildenhall2021nerf} and 3D Gaussian Splatting \cite{kerbl20233d} have demonstrated exceptional capabilities in modeling complex light transport and material interactions for photorealistic scene reconstruction. These advances in modeling optical frequencies suggest promising approaches for addressing challenges in radio frequency (RF) propagation modeling, as both domains share fundamental physical principles. RF signals, like visible light, follow Maxwell's equations and exhibit similar wave phenomena, including reflection, refraction, and diffraction. However, the much longer wavelengths, smaller antenna aperture, and ability to penetrate common objects, result in more sophisticated interaction with objects. 

Accurate modeling of RF propagation remains a challenging problem in wireless systems with diverse use cases. For instance, RF simulation enables optimal deployment of access points \cite{liu2019optimization} and intelligent reflectors \cite{ma2024automs, qian2022millimirror,fu2024adaptive},  by computing signal coverage across all possible deployment configurations. While industry-standard ray tracing algorithms \cite{wi, afc, chen2023rf} approximate electromagnetic wave behavior through ray-object interactions, they often exhibit significant simulation-to-reality gaps due to simplified modeling of surface characteristics and material electromagnetic properties. Recent neural approaches in RF modeling, such as NeRF$^2$ \cite{zhao2023nerf2}, have shown potential in learning detailed RF-object interactions. However, these data-driven models require dense spatial sampling and complete model retraining when environmental conditions change, highlighting the need for more efficient and adaptable approaches.

In this paper, we introduce \name, a novel RF simulation framework that synergistically integrates the high-fidelity of neural representation and the flexibility and interpretability of ray tracing.  
The core novelty of \name lies in its approach of generating dedicated neural representations for individual objects within a scene and seamlessly importing these representations into a ray tracing pipeline. 
By implicitly modeling the minute structural details and complex material properties of objects using neural representations, \name captures the intricacies of RF signal propagation and interactions. Simultaneously, \name 
preserves the inherent flexibility of traditional ray tracing methods, enabling editing and dynamic updates of the simulated RF environment as objects are added, moved, or removed.  To enable optimization of scene geometries in \name,  differentiable ray-geometry intersections are needed for gradient-based optimization. We choose to represent geometries using Signed Distance Functions (SDFs), as they provide continuous and differentiable distance fields that are compatible with automatic differentiation, making them suitable for gradient-based optimization of object shapes and positions \cite{Vicini2022sdf,bangaru2022differentiable,wang2021neus}.

By decoupling object representations from ray propagation modeling, \name facilitates flexible editing of an RF simulation scene, such as modification of objects and configuration of radio hardware. This flexibility is highly desirable for the design and verification of wireless communication and sensing systems. Moreover, \name paves the way towards a modularized primitive library, analogous to the well-established practice of curating libraries of 3D mesh models in computer graphics. 
Such a library would enable large-scale, modularized RF simulation through pre-trained object representations, primed for plug-and-play integration into the design process of wireless systems.

We have implemented \name and performed extensive evaluation using various wireless testbeds, including WiFi radios (both sub-6 GHz and 60 GHz bands) and millimeter-wave (mmWave) radar.
Our microbenchmark results demonstrate that \name generates RF signals that match the ground truth with high fidelity, achieving 13~dB improvement in RF-object characterization compared to conventional visual model based ray tracing. With merely 1.25~samples/sq ft of training data, \name outperforms the state-of-the-art (SOTA) NeRF$^2$ by around 5~dB in WiFi channel prediction, and even higher for mmWave. In addition, \name can adapt to changes in a scene with a small set of 3 to 5 additional data samples.

Our contributions can be summarized as follows:
\begin{itemize}

\item We propose \name, a novel RF simulation framework that reduces the sim-to-real gap of conventional ray tracing while maintaining flexibility. 

\item We design a novel neural representation that can capture detailed structure and material properties of objects while ensuring compatibility with ray tracing. We further present an efficient training strategy for learning these neural representations.

\item We evaluate \name extensively across diverse real-world wireless environments. Our results demonstrate \name's effectiveness and its potential to foster new applications of RF simulation.
\end{itemize}

\begin{figure*}[htb!]
\centering
\includegraphics[width=0.99\textwidth]{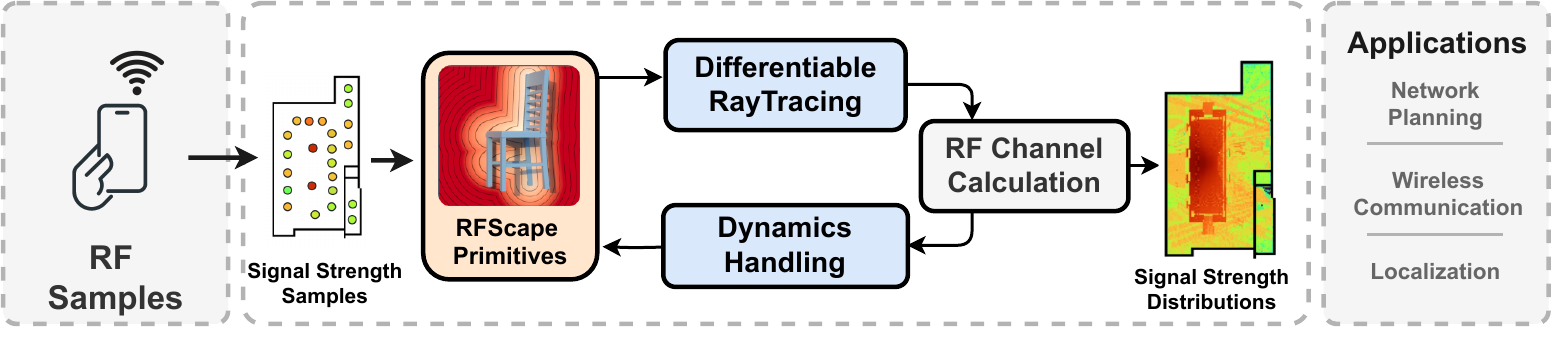}
\caption{An overview of the \name framework.}
\label{fig:framework}
\vspace{-10pt}
\end{figure*}

\section{Related Work}
\label{sec:prelim}

\noindent\textbf{RF propagation and EM field simulation.}
RF signal propagation and environmental interaction simulation is essential for wireless communication and sensing system design.
Full-wave simulation methods, including Finite Element Method (FEM) \cite{jin2015finite} and Finite-Difference Time-Domain (FDTD) \cite{taflove2005computational}, provide high-accuracy results but are computationally intensive, limiting their application to small-scale RF device design such as antenna development. 
For large-scale simulations, ray tracing methods~\cite{sbr,afc} offer computational efficiency by modeling electromagnetic waves using geometrical optics principles.
Ray tracing simulation encompasses transmitters (Tx), receivers (Rx), and mesh-based object representations within a scene. 
The mesh representation comprises multiple polygonal elements that approximate the object's surface geometry.
The simulation initiates with ray emission from each Tx across all directions, where angular resolution parameters define the ray density distribution. 
The simulator subsequently computes ray-mesh intersections and, at each intersection point, generates secondary rays based on the number of specified interactions.
These secondary rays are propagated according to their interaction mechanisms (e.g., reflection, refraction, or diffraction). The simulator also computes the \textit{directional radiance}, which quantifies the ray's power distribution along specific directions post-interaction, determined by the material properties (e.g., permittivity and conductivity) of the intersected objects.
Ray propagation terminates upon reaching either the Rx or a specified maximum interaction threshold.
The simulator aggregates all received rays at the Rx through coherent combination, computing the relative phase differences based on their respective propagation path lengths.

The accuracy of ray tracing simulations depends critically on precise object geometry and material property models, which exceed the capabilities of conventional sensing systems. Notable challenges arise even for specular objects with simple geometric structures, as visual 3D reconstruction methods often fail to capture microscale surface features that, while visually insignificant, can substantially influence RF signal interactions.

\noindent\textbf{Neural representations for EM simulation.}
Neural scene representations have established a significant position in computer vision and computer graphics, particularly following the introduction of Neural Radiance Fields (NeRF) \cite{nerf}.
NeRF implements neural networks to model two fundamental scene properties implicitly: the volumetric density at each spatial coordinate and the directional radiance (comprising color and intensity) emitted from that coordinate.
Through learning these spatially and directionally dependent radiance functions from multiple input images, NeRF generates photorealistic novel viewpoints without explicit geometric or surface representations. 

There have been efforts to adapt neural representation for RF applications \cite{NeWRF,chen2024rfcanvas}, with NeRF$^2$\cite{zhao2023nerf2} being a prominent example. This method models the entire scene as an implicit neural representation, preventing the clear delineation of object boundaries and their RF interaction characteristics. This limitation introduces constraints on system adaptability and increases susceptibility to environmental variations. While WiNeRT~\cite{orekondy2022winert} advances this approach by incorporating neural network-based material reflection parameters into ray tracing simulations, its reliance on traditional 3D mesh representations remains insufficient for capturing detailed structural features of physical objects \cite{seyb2024microfacets}.

\name employs a learnable geometry representation with a neural material network to capture the complex and intrinsic RF properties of physical objects. Additionally, \name separates empty space and medium to reduce optimization complexity and improve sampling efficiency.

\section{Preliminaries }

\subsection{RF Ray Tracing }
The proliferation of wireless communications and sensing systems has made accurate modeling of RF propagation increasingly critical. While full-wave electromagnetic (EM) solutions provide high accuracy, their computational complexity makes them impractical for large-scale scenarios. RF ray tracing offers an efficient alternative by applying geometric optics principles to electromagnetic wave propagation, enabling fast yet accurate predictions of radio coverage in complex environments. RF waves interact with the environment through reflection, refraction, diffraction, and scattering. These interactions are governed by Maxwell's equations but can be approximated using ray optics when the wavelength is much smaller than environmental features.

The workflow begins with environment modeling, where the physical space is discretized into polygonal surfaces with assigned electromagnetic properties. Each material $i$ is characterized by its complex permittivity $\varepsilon_i$ and conductivity $\sigma_i$:

\begin{equation}
\vspace{-5pt}
    \varepsilon_c = \varepsilon_i - j\frac{\sigma_i}{\omega\varepsilon_0},
\end{equation}

\noindent where $\omega$ represents the angular frequency and $\varepsilon_0$ is the free space permittivity. From each transmitter location $\mathbf{r}_t$, rays are launched at discrete angles $(\theta, \phi)$ and traced according to Fermat's principle:

\begin{equation}
    \frac{d}{ds}\left(n\frac{d\mathbf{r}}{ds}\right) = \nabla n,
\end{equation}

\noindent where $n$ denotes the refractive index of the medium, $s$ represents the path length parameter, and $\mathbf{r}$ is the position vector. 

\subsubsection{RF Channel Calculation}
The electric field $\mathbf{E}$ at receiver position $\mathbf{r}_r$ is computed by summing contributions from all paths:

\begin{equation}
    \mathbf{E}(\mathbf{r}_r) = \sum_{p=1}^{N_p} A_p e^{-jk_0s_p} \prod_{k=1}^{N_i} \mathbf{R}_k \cdot \mathbf{E}_0,
\end{equation}

\noindent where the amplitude spreading factor $A_p$ accounts for geometric attenuation, $k_0$ represents the free space wavenumber, $s_p$ denotes the total path length, $\prod_{k=1}^{N_i} \mathbf{R}_k$ incorporates all reflection and transmission coefficients along the path, and $\mathbf{E}_0$ represents the reference field at unit distance. The reflection coefficient for perpendicular polarization follows:

\begin{equation}
    R_\perp = \frac{\eta_2\cos\theta_i - \eta_1\cos\theta_t}{\eta_2\cos\theta_i + \eta_1\cos\theta_t},
\end{equation}

\noindent where $\eta_i$ represents the wave impedance in medium $i$, and $\theta_i$, $\theta_t$ are the incident and transmitted angles respectively.

\subsection{Signed Distance Field (SDF)}
Inspired by recent advances in computer graphics~\cite{Vicini2022sdf,iron-2022}, we propose to represent the geometric structure of an object using an SDF. Given an object \(\Omega\) and its boundary \(\partial \Omega\), an SDF \(S_\Omega(p)\) defines the signed distance between any point $p$ in the space and the object's surface \(\partial \Omega\):
\begin{equation}
S_{\Omega}(p) = s(p,\Omega)\cdot d(p, \partial \Omega),
\end{equation}
where $d(p, \partial \Omega)=\inf_{q \in \partial \Omega} d(p, q)$ denotes the minimal distance between the point $p$ and any point $q$ on the object's surface $\partial \Omega$. In addition, the sign $s(p,\Omega)$ indicates the \textit{containing relation} between the point $p$ and the object $\Omega$, where a positive sign means that the point \(p\) is outside the object, \ie, \(p \notin \Omega\) while a negative sign means \(p \in \Omega\). 
Intuitively, the SDF clearly defines the object's boundary, which are the solutions to the equation $S_\Omega(p)=0$.

One fundamental property of SDF is the satisfaction of the Eikonal equation~\cite{sethian1996fast}:
\begin{equation}
\label{eq:eikonal}
\|\nabla S_{\Omega}(p)\| = 1.
\end{equation}
It implies that the gradient of \(S_{\Omega}\) at any point on the surface \(\partial \Omega\) has a unit magnitude. Therefore, $\nabla S_{\Omega}(p)$ is exactly the normal vector $n$ of the surface, which is necessary information for ray tracing to calculate new directions of RF rays after interacting with an object.
In practice, the normal vector \(n(p)\) at any point \(p\) on the surface can be approximated by finite differences: 
\begin{equation}
\label{eq:fd_norm}
n(p) = \nabla S_{\Omega}(p) = \begin{bmatrix}
\frac{S_{\Omega}(p+\epsilon_x)-S_{\Omega}(p-\epsilon_x)}{2\epsilon_x} \\ 
\frac{S_{\Omega}(p+\epsilon_y)-S_{\Omega}(p-\epsilon_y)}{2\epsilon_y} \\ 
\frac{S_{\Omega}(p+\epsilon_z)-S_{\Omega}(p-\epsilon_z)}{2\epsilon_z}
\end{bmatrix},
\end{equation}
where \(\epsilon_x\), \(\epsilon_y\), and \(\epsilon_z\) are small perturbations in the \(x\), \(y\), and \(z\) directions, respectively, and \(\epsilon\) is the magnitude of these perturbations.

Following the above definition, every object occupying physical space has an associated SDF that implicitly encodes the object's precise surface details. In addition, SDF defines a continuous signed distance field rather than discrete polygon surfaces, thus enabling gradient-based end-to-end optimization.
However, SDF itself does not encode any material information, and lacks explicit definition of the object boundaries -- both critical for RF ray tracing.

\begin{figure*}[htb!]
\centering
\includegraphics[width=0.99\textwidth]{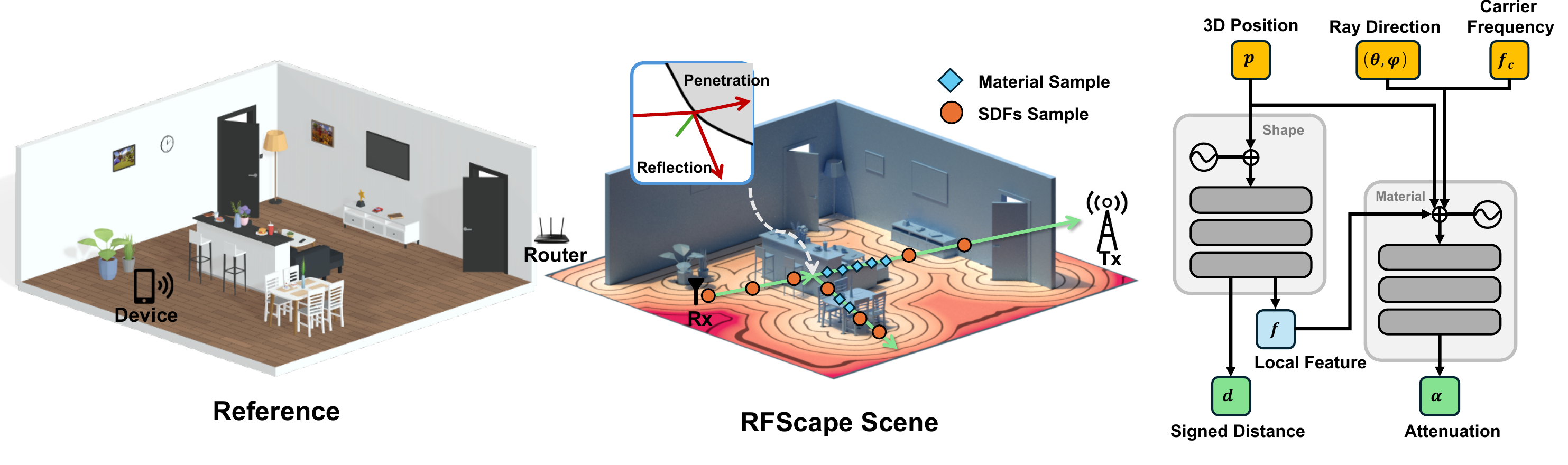}
\caption{RF Ray Tracing with \name's Neural Object Representations. While electromagnetic waves physically propagate from Tx (router) to Rx (device), the ray tracing computation traces paths backward from Rx to Tx for efficiency. Objects along these paths are modeled by our learned neural representations.}
\label{fig:framework2}
\vspace{-10pt}
\end{figure*} 

\section{Method}

\name is an RF simulation framework that integrates high-fidelity object models into a ray tracing engine, achieving both high accuracy and flexible editability for complex and even dynamic scenes. The framework consists of three primary components: 1) \textit{\name primitives}, which model the geometric structures and material properties of objects for RF simulation; 2) \textit{RF Ray Tracing}, which accurately simulates RF interactions; and 3) \textit{Dynamics Handling}, which incorporates prior knowledge about scene changes to modify the \name scene. Additionally, \name's optimization module enables end-to-end differentiable forward simulation and gradient-based backward optimization of the primitives' parameters.

\subsection{\name Primitives}

A \name primitive represents an object in a scene to be simulated via ray tracing. Such a primitive must meet two requirements: 1) encoding of fine-grained geometric structure and material properties of the object, and 2) seamless interaction with RF rays. The first requirement can be fulfilled using trainable neural representations, such as NeRF \cite{nerf,zhao2023nerf2}. To further meet the second requirement, we need 
the neural representation to clearly define the boundary of the object, so that the geometric intersections between RF rays and objects can be localized.
Unfortunately, as a scene-level model, NeRF obscures the clear delineation of individual objects. On the other hand, conventional mesh-based geometric representation suffers from inaccuracy (Sec.~\ref{sec:prelim}) and discontinuity, impeding its integration into an end-to-end optimization model.

\subsubsection{Neural Representations of Geometric Structure and Material}
To approximate the geometric structure of the object with fine-grained details, we propose to 
represent the SDF of the object using a \textit{structure model}. 
As shown in Fig.~\ref{fig:framework2}, our structure model is a trainable neural network, which takes as input the position $p$ of a    
marching ray and outputs a signed distance, subsequently used for sphere ray tracing.
By learning the neural SDF representation, \name can eventually localize the interactions between rays and objects with infinitesimally small errors, potentially eliminating the sim-to-real discrepancies that plague conventional mesh representation.  

However, only knowing the intersection of the ray and the object is insufficient for calculating the reflective and penetrative rays, 
which additionally depends on the local material properties of the object. 
Conventional RF ray tracing \cite{sbr} usually calculates the Fresnel coefficients, assuming that the object has known and homogeneous material properties. 
Unfortunately, a practical object's material properties are distributed unevenly across its surface and interior and are often not measurable. To overcome this challenge, we employ a second neural network to model the object's material distribution and learn the 
directional attenuation coefficients caused by materials.
As shown in Fig.~\ref{fig:framework}, the inputs of this material model include any position $p$ of the marching ray, the carrier frequency $f_c$ of the RF signal, the local structure features $f$ generated by the structure model, and the direction $(\theta, \phi)$ of the resulting reflective or penetrative rays. The output is an attenuation coefficient $\alpha$ along that direction. 

We emphasize that this model encapsulates all RF propagation effects (reflection, penetration, scattering, \etc) through the directional attenuation coefficient. These effects can all be represented by the attenuation along arbitrary directions after a marching ray interacts with a point on/within the object. 
In addition, unlike conventional ray tracing, where interactions only happen at the boundaries of objects, our material model accounts for the \textit{interior} of the object.
Specifically, when the position of a marching ray is inside the object, \ie, $p\in\Omega$, the material model can output the attenuation coefficient for the propagation direction of the ray.

Following the success of the NeRF model \cite{nerf}, we implement both structure and material models as multi-layer perceptrons (MLPs). 
The interaction between a marching ray at $p$ and the \name primitive of an object $\Omega$ can be formalized as follows: 
\begin{equation}
\text{M}_{\Omega} : (p,\theta,\phi,f_c) \rightarrow (d,\alpha).
\end{equation}
Both the signed distance $d$ and the directional attenuation coefficient $\alpha$ are used to update the ray upon any interaction, including penetration and reflection along all possible directions.

\subsection{ RF Ray Tracing with \name Primitives}
Signal propagation along each ray path involves complex interactions with one or multiple objects. A vital step in ray tracing is to locate the interaction. Conventional ray tracing algorithms widely adopt the Möller–Trumbore algorithm~\cite{moller2005fast} to compute the intersections between the rays and the polygons of a mesh-based object model. 
In contrast, \textit{SDFs represent surfaces through continuous values without an explicit object boundary}. To overcome this challenge, \name locates the intersection point \(p\) by iteratively searching for the SDF zero-crossing along the ray direction. The iterative process is as follows:

\begin{equation}
T(p_0,\omega_0) = \sum_{i=0}^{N} S_{\Omega}(p_i), \quad p_{i+1} = p_i + \omega_0 \cdot S(p_i), 
\end{equation}
\noindent, where $p_0$ and $\omega_0$ denote the initial position and propagation direction of the ray, and $T$, is the total distance traveled by the ray. The search terminates when $|S(p_i)| < \epsilon$, where $\epsilon=0.01$, or when it reaches maximum iteration depths $N$. This is known as \textit{sphere tracing}~\cite{hart1996sphere} for SDF-based object models.

Upon intersection, \ie, $d=0$, \name estimates the local normal vector $n$ using Eq.~\eqref{eq:fd_norm} and determines the direction of the reflection ray, \ie, $\omega_{r}=\omega_0 - 2(\omega_0 \cdot n)n$ according to Snell's Law, and the direction of the penetrative ray, \ie, $\omega_{t}=\omega_0$.

Given the directions of the rays after interactions, we can use the material model to obtain the attenuation coefficients for reflection and penetration, \ie, $a_{r}$ and $a_{t}$. To constrain the complexity of the \name ray tracing, we follow a Monte Carlo strategy and only select either of 
the reflective or penetrative ray for further propagation, with probabilities proportional to their directional attenuation coefficients.

Penetrating through an object, \ie, $d<0$, is similar to propagation in the free space, except that the attenuation coefficients are estimated using the material model and accumulated to obtain the total attenuation within the object, \ie, $a_{\Omega} = \int_{t_{\text{in}}}^{t_{\text{out}}} M(p_0 + t \cdot \omega) \, dt$. 
When the interior ray reaches the boundary of the object again, the same Monte Carlo strategy is used to select either the internally reflective ray or the exiting ray.

Following the conventional ray tracing, \name emits rays from a Tx, calculating all interactions between the rays and the objects in the scene and eventually identifying the rays that reach the Rx to establish valid paths. Each valid path is characterized by its traversal distance \(\tau\) and 
attenuation coefficient. We record all the intermediate segments of the rays and calculate the total distance $ \tau$ and attention coefficient $a$ along the path:
\begin{equation}
\tau = \sum_i t_i, \quad  a = \prod_i a_i,
\end{equation}
and the received signal can be calculated as:

\vspace{-2mm}
\begin{align}
\label{eq:ray_tot}
S_{TX} = a \cdot S_{RX}   \cdot e^{-j2\pi f_c \tau}. 
\end{align}
\vspace{-2mm}

\subsection{\name Optimization}
\label{subsec:optimization}
To capture the real structure and material properties of the objects and accurately represent the scene with \name primitives, we need to optimize their parameters $\Theta$ to minimize the sim-to-real discrepancy, \ie, 
\begin{equation}
\label{eq:obj_func}
\Theta^* =  \underset{\Theta}{\text{arg} \textit{min}}  \sum_{i=0}^M \mathcal{L}(\hat{s}_i,s_i), 
\end{equation}
where $\hat{s}$ are RF measurements in the scene and $M$ is the number of measurements. $\mathcal{L}$ is the loss that quantifies the discrepancy between real measurements $\hat{s}$ and \name's RF simulation output $s$. 
We expect that a few shots of RF measurements, \ie, a small $M$, is sufficient since \name obviates the need to capture expansive empty regions and only focuses on objects in the scene. Besides, the physical laws of ray propagation pose extra constraints on the \name primitives of objects.

To promote the generation of continuous surfaces by the SDFs and ensure adherence to the eikonal constraints in Eq.~\eqref{eq:eikonal}, 
we further apply discrete Laplacian and eikonal regularization. This guarantees that the norm of the gradient remains unity, maintaining the physical plausibility of the neural SDF models.

\subsection{Handling Scene Dynamics}
\label{sec:edit}
Owing to the compatibility with conventional ray tracing,
\name can support diverse radio configurations and dynamic scenes with little or even zero retraining efforts. 

\subsubsection{Tx/Rx Configurations}
\name supports flexible configurations of radio antennas in simulation. Ideally, rays are shot uniformly in all directions with equal strength at Tx, while rays arriving at Rx are equally combined, assuming both Tx and Rx antennas are isotropic. For a practical antenna with known directivity (\ie, antenna gain pattern), \name applies weights that are proportional to the antenna's directive gains to rays along different directions.
For a phased array that consists of multiple antenna elements, we can view the array as an equivalent directional antenna with switchable beam patterns. The radiance field of the antenna beam patterns can thus be simulated separately. 
By incorporating these antenna configurations into the ray tracing algorithm, we can directly reuse the \name primitives to simulate the performance of different radio configurations in a scene.

\subsubsection{Object Dynamics}
\label{sec:scene_dynamics}
\name can handle dynamic scenes where objects can be added, removed, or moved.
We assume that the \name primitives of newly added objects are available in a pre-trained \name library. 
Otherwise, they can be trained following the optimization procedure in Sec.~\ref{subsec:optimization}, by freezing the already trained \name primitives of other objects in the scene.
Unlike scene-wise neural representations \cite{zhao2023nerf2,NeWRF} that have to be completely retrained even upon partial environmental changes such as object locations, \name only needs to adapt to the partial changes, depending on how these changes are specified.

First, when visual sensors (\eg, cameras or lidars) are available, \name can utilize them to extract the identities of moving objects and their poses (\ie, translation and rotation vectors) with sufficient accuracy. 
Given the updated pose of an object, \name transforms the global position and directions of marching rays into the original coordinate system of this object before it moves, so that its \name primitive can be used to model the interactions with the rays.
This allows for updates of the scene based on the visual changes in the environment.

Second, when visual sensors are unavailable but RF measurements (\eg, signal strength) can be sampled within the scene, we can 
\textit{freeze} the \name primitives of the objects, and only optimize the poses of the foreground dynamic objects following Sec.~\ref{subsec:optimization}. 
As inputs for the optimization, the RF measurements can be collected by devices already deployed within the scene.

We note that similar to conventional ray tracing applications,  it is also common that the scene dynamics are specified by users as part of an RF simulation process, \ie, when specifying and iterating over possible base station locations to optimize coverage.  

\section{Experiment}

\subsection{Experimental Settings}
\label{sec:setup}

\noindent\textbf{Dataset} 
We conduct real-world wireless measurements using a mobile robotic platform. We utilize a Turtlebot4 equipped with an integrated LiDAR sensor and Simultaneous Localization and Mapping (SLAM) system for precise navigation and positioning. The platform follows predefined trajectories with designated waypoints to ensure consistent data collection across the target space. For wireless signal acquisition, we deploy an ASUS RT-AC86U router operating in concurrent dual-band mode (2.4 GHz and 5 GHz) as the access point, paired with an iPhone 14 Pro client device leveraging the AirPort Utility application \cite{Apple_2011} for Received Signal Strength Indicator (RSSI) measurements. Additionally, our setup incorporates two 802.11ad-compliant MikroTik wAP 60G$\times$3 routers \cite{blanco2022augmenting} for 60~GHz WiGig measurements, running OpenWrt with Mikrotik Researcher Tools \cite{IMDEA} to enable RSSI collection.

\begin{figure}[htb!]
\centering
\includegraphics[width=0.48\textwidth]{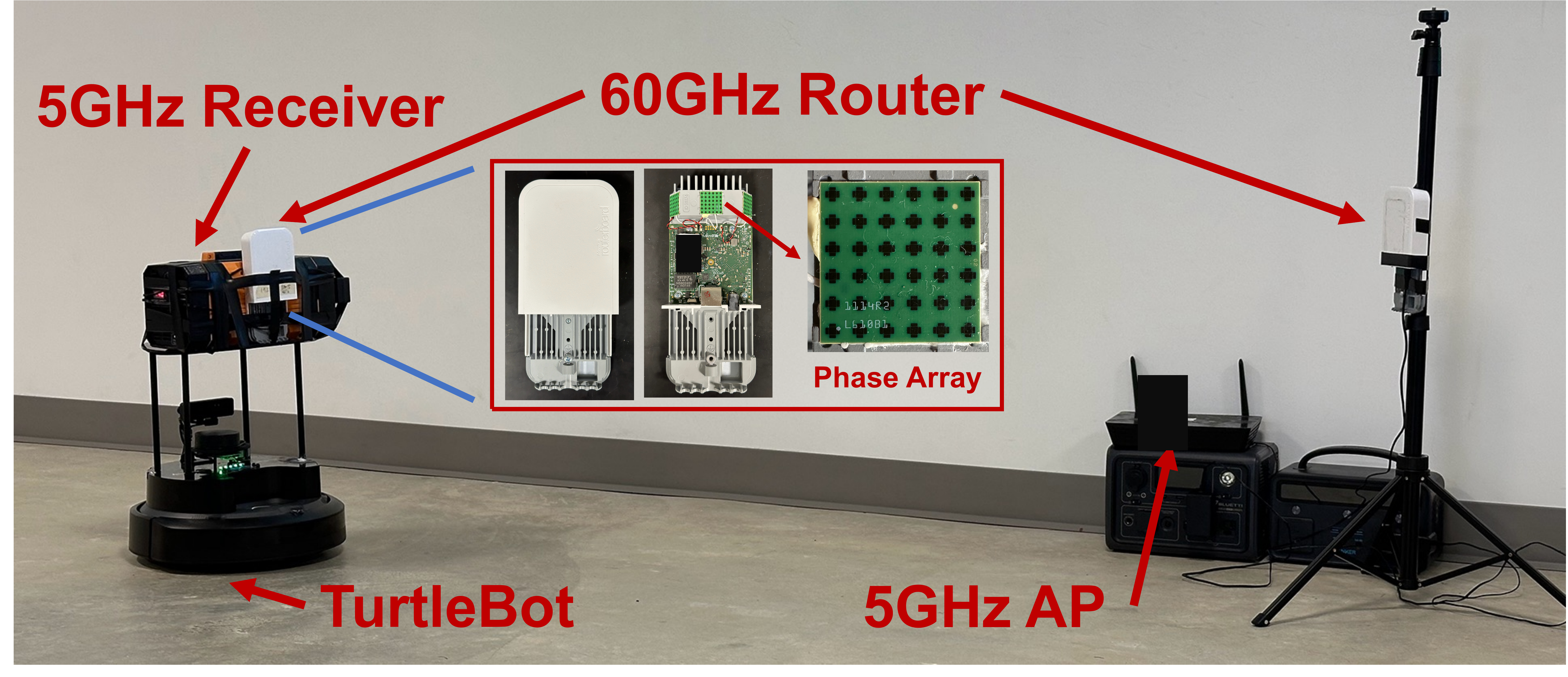}
\caption{Room-level Wireless Channel Distribution Collection Setup. }
\label{fig:setup}
\end{figure}
\begin{figure}[htb!]
    \centering
    \includegraphics[width=1\linewidth]{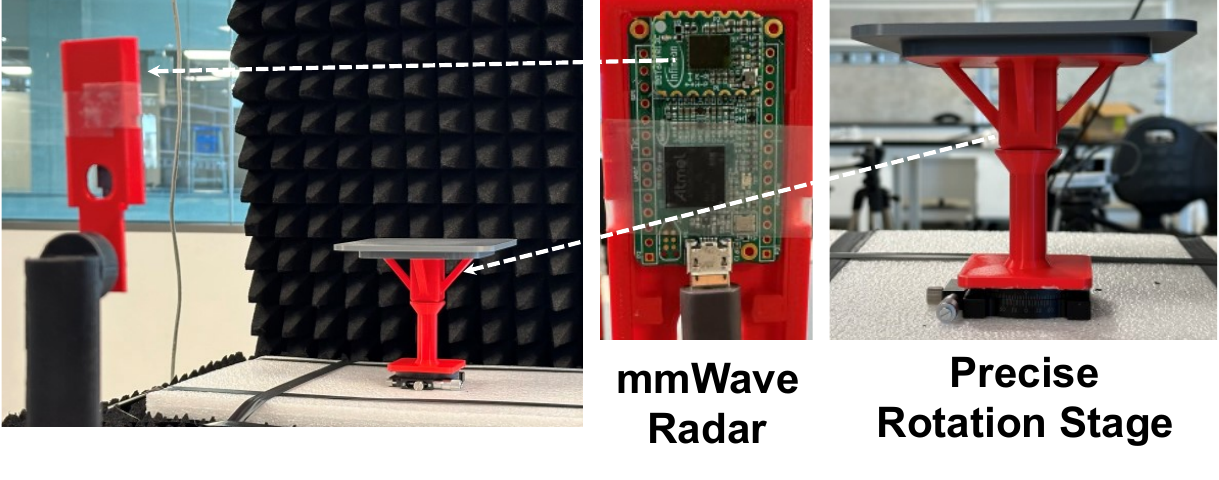}
    \caption{Object-level Wireless Channel Response Collection Setup.}
    \label{fig:setup_rotate}
\end{figure}

\noindent\textbf{Baselines}
We employ NeRF$^2$ \cite{zhao2023nerf2} as the baseline, as it represents the state-of-the-art in neural channel prediction, outperforming other approaches such as Deep Convolutional Generative Adversarial Network (DCGAN)\cite{radford2015unsupervised} and Variational Autoencoder (VAE)\cite{kingma2013auto}. Our implementation follows the default configuration provided in the publicly available code archive. The reproduced results show a performance of 2.37 dB on BLE RSSI, 22.24 dB on WiFi SNR, and 0.82 SSIM for the RFID spectrum, which aligns with the original paper.

\noindent\textbf{Model Implementation and Training}

\begin{itemize}
    \item \textbf{Structure model:} 
    We employ an 8-layer Multilayer Perceptron (MLP) with a width of 64 and integration of skip connections. To address the spectral bias inherent in MLPs, the input 3D location $x$ undergoes positional encoding with 6 frequencies. The actual implementation is $S_{\Omega}: \ x \rightarrow (d,f)$, where we output additional 128-dimensional local features $f$ for the material model.

    \item \textbf{Material model:} 
    The material model consists of an 8-layer MLP with a width of 256 and incorporates skip connections. The spatial inputs ($x$, $n$, $\theta$) enhance the network's sensitivity to spatial variations. This architecture is designed to encapsulate the intricate RF-material interactions, ensuring a faithful representation of material properties and electromagnetic wave behavior.
\end{itemize}

The training configuration includes a batch size of 4096 and utilizes the Adam optimizer~\cite{kingma2014adam}. The learning rate is initialized at $3 \times 10^{-4}$ and gradually decays to $3 \times 10^{-5}$ following an exponential schedule. Default values are maintained for other hyperparameters, such as $\beta_1 = 0.9$, $\beta_2 = 0.999$, and $\epsilon = 10^{-7}$. On an NVIDIA A6000 GPU, the network typically converges after approximately $300k$ iterations when training on a single scene.

\subsection{Results}
In this section, we conduct a microbenchmark experiment to evaluate \name's capability in characterizing the interaction between RF signals and objects.  
We place individual objects on a precise rotation stage with RF absorbers around to isolate ambient multipath, as shown in Fig.~\ref{fig:setup_rotate}. 
We position an FMCW radar 1m from the target objects to transmit mmWave signals and record the return signals. We rotate the objects at intervals of $5$ or $10$ degrees while recording the radar's Range-FFT signals. We randomly select $50\%$ of the collected data for training and reserve the remaining $50\%$ for testing, ensuring uniform distribution across rotation angles. The evaluation includes three objects with increasing geometric complexity: a kettle, a teacup, and a robot.

\subsubsection{ Object-level RF Characterization }
\begin{figure}[htb!]
\centering
\includegraphics[width=0.48\textwidth]{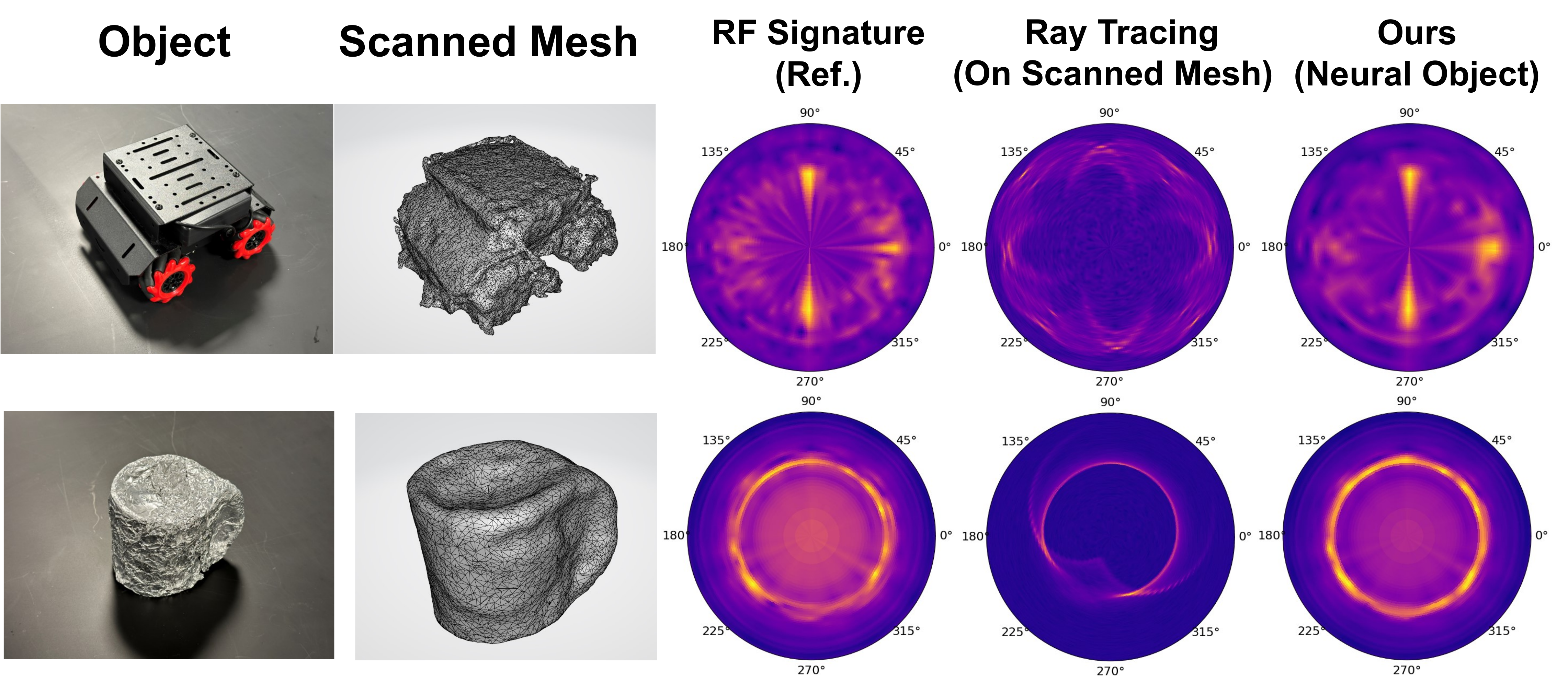}
\caption{RF Characterization Prediction Results. }
\label{fig:char}
\vspace{-10pt}
\end{figure} 

To establish a baseline, we employ Polycam \cite{polycam-website}, a highly optimized commercial 3D mesh scanner application \cite{wang2023pointshopar}, to capture mesh models of the objects. These models are then simulated using well-established ray tracing algorithms. The material parameters used in the simulations are obtained from relevant literature \cite{mat1,mat2}.

The results are shown in Fig.~\ref{fig:char}.  The mesh models obtained from visual scans fail to capture accurate microgeometry. They also assume objects have homogeneous materials throughout. Consequently, directly applying ray tracing on visually scanned meshes yields significant errors, with a large median error of 15.7~dB. In contrast, \name leverages neural object representations to model the RF signatures of objects using only 
around one sample/sq ft of RF observations, reducing the median error to just 2.9~dB while still preserving the scene editability. This makes \name a much more accurate option for RF object modeling compared to the mesh-based approaches commonly used in traditional ray tracing. It is worth noting that NeRF$^2$ did not demonstrate learning from FMCW radar signals, and the objects fit by NeRF$^2$  cannot be edited or re-simulated, hence its exclusion from this experiment.

\begin{figure*}[ht!]
 \begin{minipage}{0.33\linewidth}
        \setlength{\abovecaptionskip}{0pt}
        \centering
        \includegraphics[width=1\linewidth]{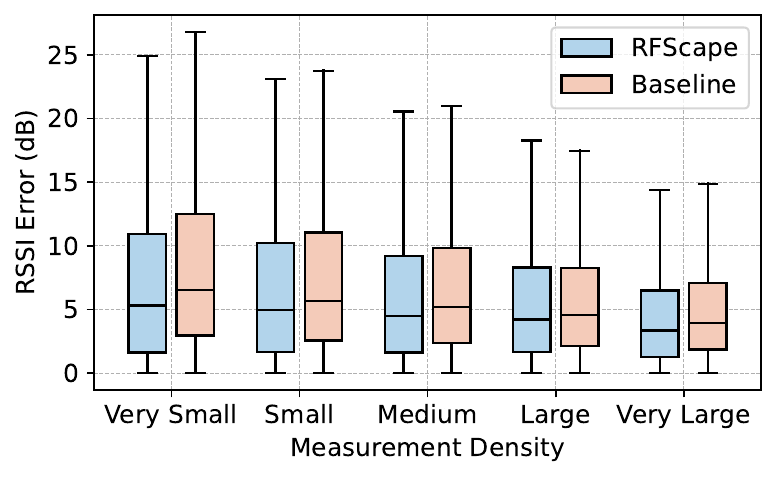}
        \caption{BLE RSSI (Lower values indicate better performance) }
        \label{fig:ble}
    \end{minipage}
    \begin{minipage}{0.33\linewidth}
        \setlength{\abovecaptionskip}{0pt}
        \centering
        \includegraphics[width=1\linewidth]{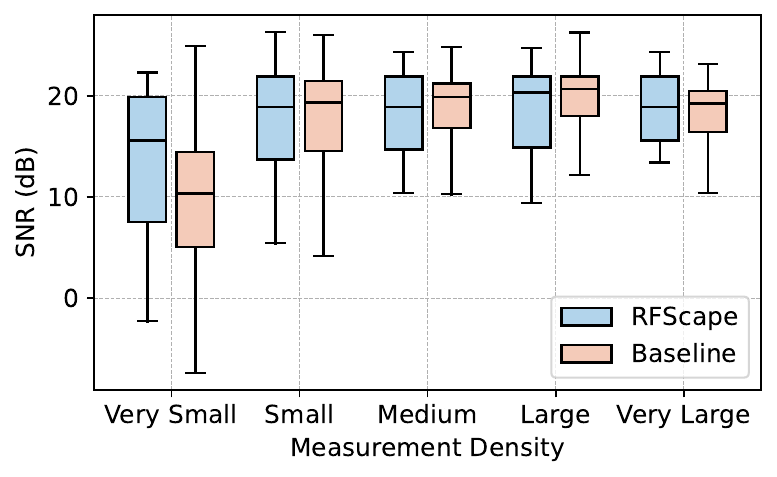}
        \caption{2.4 GHz WIFI CSI (Higher values indicate better performance) }
        \label{fig:wifi}
    \end{minipage}
    \begin{minipage}{0.31\linewidth}
        \setlength{\abovecaptionskip}{0pt}
        \centering
        \includegraphics[width=1\linewidth]{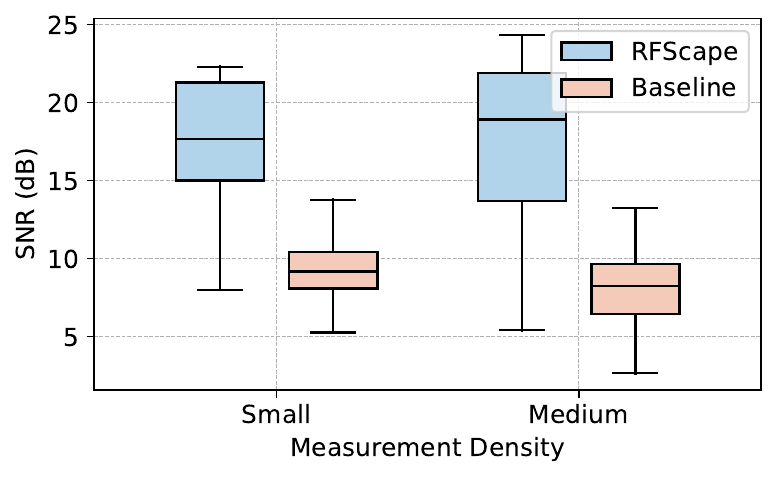}
        \caption{60GHz WiGig CSI   (Higher values indicate better performance) }
        \label{fig:wigig}
    \end{minipage}
    \vspace{-10pt}
\end{figure*}

\subsubsection{Room-Level RF Channel Prediction Evaluation}

To evaluate the performance and scalability of \name in predicting the RF channel distribution, we compare it with the state-of-the-art NeRF$^2$\cite{zhao2023nerf2} framework. The experiments are conducted with 5-fold cross-validation using the dataset provided by NeRF$^2$, which includes 2.4 GHz WiFi CSI, BLE RSSI, and RFID spectrum data \cite{zhao2023nerf2}, as well as our self-collected 60GHz WiGig CSI and additional 5GHz WiFi data (Sec.~\ref{sec:setup}). NeRF$^2$ requires approximately 200 channel measurements per square foot as training data, which presents significant limitations in practical applications. To evaluate performance under realistic conditions, we evaluate both \name and NeRF$^2$ across multiple training data densities: very small (0.625 samples/sq ft), small (1.25 samples/sq ft), medium (2.5 samples/sq ft), large (5 samples/sq ft), and very large (10 samples/sq ft).

Fig.\ref{fig:ble}, \ref{fig:wifi}, and  \ref{fig:wigig}  present the comparison results between \name and NeRF$^2$ under different wireless protocols. For the small-sized WiFi CSI training dataset, \name achieves a median SNR of 15.6~dB, outperforming NeRF$^2$ by 5.3~dB. Similarly, on the small-sized RFID spectrum datasets, \name is comparable to NeRF$^2$, with SSIMs of 0.4498 and 0.5061, respectively. On the small-sized BLE RSSI datasets, \name reduces the RSSI error to 5.3~dB, which is 1.2~dB better than the baseline.

These results suggest that \textit{\name is more efficient in learning and generalizing from limited data samples}. This is due to better physical constraints and clear object boundaries that eliminate the need to capture expansive empty regions.

The advantage of \name becomes even more pronounced at higher frequency bands, as demonstrated in Fig.~\ref{fig:wigig}. The narrower beams, shorter wavelengths (hence less diffraction), and higher path loss of the mmWave signals together result in sparser signal propagation compared to lower frequency bands. Accurately predicting the mmWave channel requires finer-grained physical simulation, which is incorporated in \name. In contrast, the baseline NeRF$^2$ struggles to predict mmWave band CSI, resulting in a low SNR of 9.08~dB. \name, on the other hand, maintains a 
16.2~dB SNR in CSI channel prediction for the mmWave band. These results indicate that \textit{\name yields superior performance on high-frequency channels which are more sensitive to ray-object interactions}. 

\begin{figure}[htb!]
\centering
\includegraphics[width=\linewidth]{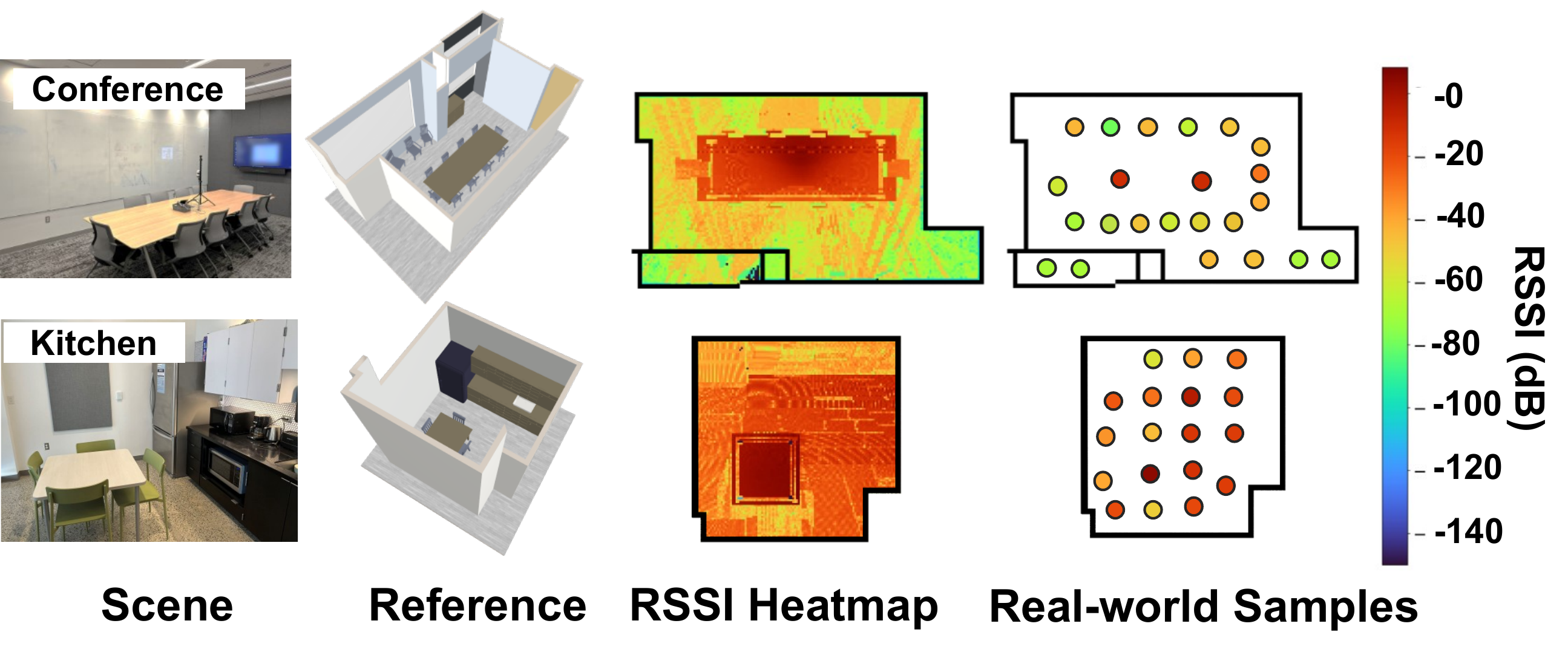}
\caption{Channel Prediction Results. }
\label{fig:room}
\vspace{-10pt}
\end{figure} 

\subsubsection{Dynamics Adaptation Evaluation}

In this section, we evaluate the adaptability of \name to environmental alterations.  The original scenario consists of a long wooden table with chairs positioned along its length, and additional plastic chairs are located in the corner of the room. We consider a scenario where, after a conference discussion, the scene is modified by introducing a blackboard and repositioning several chairs. 

To facilitate \name's adaptation to the modified environment, we evaluate the two approaches proposed in Sec. \ref{sec:scene_dynamics}. The first method involves utilizing assertions based on visual information. We assume that information regarding the location, orientation, and type of objects that are added, changed, or removed is obtained from a camera in the room.
The second approach involves \textit{recollecting a small set of 3 to 5 data points} from the modified scenario.

Subsequently, we re-scan the environment using the setup shown in Fig.\ref{fig:setup} as ground truth.  Both methods successfully update \name's scene representation and accurately predict the updated scene channels, with median RSSI errors of 2.9 and 3.2~dB, respectively. These minimal discrepancies between the adapted \name and the ground truth demonstrate \name's adaptability to scene dynamics with minimal assisted information.


\section{Conclusion}
We have presented \name, a novel framework that 
seamlessly integrates high-fidelity neural object representations with ray tracing for accurate and flexible RF simulations. 
\name is inspired by state-of-the-art SDF-based object modeling techniques in computer graphics, but overcomes their incompatibility with ray tracing and lack of models for RF-material interaction.  
The modular nature of \name enables reconfiguration of the RF simulation scenes with little or zero site-specific retraining.  As an advancement to the physics-based ray tracing simulators widely used in the wireless industry, we believe \name can serve as a powerful computer-aided design tool for diverse wireless applications, such as network planning and inverse simulation for 3D reconstruction. We leave the exploration of such applications for future work.


\newpage
{
    \small
    \bibliographystyle{ieeenat_fullname}
    \bibliography{main}
}

\end{document}